\documentclass{aa501}
\usepackage[dvips]{graphics}
\begin{document}

%
   \title{Variability in the extreme helium star 
\object{LSS 5121}\thanks{Based on observations made at the University of Texas
McDonald Observatory, Fort Davis, Texas and the Jacobus Kapteyn Telescope
operated on the island of La Palma by the Isaac Newton Group in the Spanish
Observatorio del Roque de los Muchachos of the Instituto de Astrofisica de
Canarias. }}

   \author{V. M. Woolf\inst{1}, R. Aznar Cuadrado\inst{1},
       G. Pandey\inst{2}, \and C. S. Jeffery\inst{1}
          }

   \offprints{V. M. Woolf}

   \institute{Armagh Observatory, College Hill, Armagh BT61~9DG,
                Northern Ireland \\
              email: vmw@star.arm.ac.uk, rea@star.arm.ac.uk, csj@star.arm.ac.uk
     \and
              Department of Astronomy, University of Texas,
              Austin, TX 78712, USA \\
	      email: pandey@astro.as.utexas.edu
             }

   \date{}

   \titlerunning{LSS 5121}

   \abstract{
We report a photometric and spectroscopic study of the hot extreme helium
star \object{LSS 5121}.  We found photometric variability, but no
period was evident in its periodogram.  This is consistent with the previous
proposal, based on spectral line variations, that LSS~5121 is a non-radial
pulsator similar to other hot extreme helium stars.
}

\maketitle

%

\section{Introduction}

Extreme helium stars (EHes) are luminous blue stars with spectra that indicate
surfaces composed almost entirely of helium, with traces of carbon and nitrogen
and very little (usually $\ll 1$ per cent) hydrogen.  Their temperatures fall in
the range ${\rm 8000 K \leq T \leq 32\,000 K}$.
The EHes are of interest because, among other things, some of
them pulsate or are otherwise variable, they have very unusual chemical
compositions, and they are quickly evolving, with the radii and temperature
changes of some being measurable over a period of a decade or less (Jeffery
et~al. \cite{jshp01}).  Although
their evolutionary history is still not completely certain, it is evident that
they are post AGB stars evolving to become white dwarfs.  EHes evolution may
be linked with other hydrogen deficient star classes such as RCrB stars and
helium subdwarf B stars (He sdBs).
It may be that a sizable fraction of stars go through an EHe phase.

Pulsations in EHes are excited through iron group (Z-bump)
opacities (Saio \cite{s93}).
The pulsations in some EHes allow accurate estimates to be made
of their physical properties, including their mass and radii (Lynas-Gray et~al.
\cite{lsh84}, \cite{lks87}, Woolf \& Jeffery \cite{wj00}).  The study of
pulsations in stars with atmospheres not dominated by hydrogen (e.g. EHes and
He sdBs) provides additional constraints on pulsation theory.  Fewer than 30
EHes are known (Jeffery et~al. \cite{jhh96}) and few of those
have been studied in any detail.
The physical properties and the variability of additional EHes must
be determined if a statistical study of the group is to have meaning.

\object{LSS 5121} is a hot extreme helium star.
The temperature of \object{LSS 5121} has been estimated as 28\,300~K (Heber
et~al. \cite{hjd86}) using UBV color measurements and  29\,772~K (Jeffery
et~al. \cite{jshp01}) by comparing optical photometry and IUE
ultraviolet spectra with the calculated theoretical flux distributions.
Earlier spectroscopic observations by Lawson and Kilkenny (\cite{lk98})
demonstrated that its absorption lines vary in strength on a timescale of a
few days.  The line variations are similar to those observed in the
(${\rm T \approx 20~kK}$) EHe \object{BD$-9\degr 4395$}, a 
non-radial pulsator (Jeffery \& Heber \cite{jh92}).
The spectrum of \object{LSS 5121} is very similar to that of the
(${\rm T \approx 29~kK}$) EHe \object{HD 160641} (Heber et~al. \cite{hjd86}),
which has been found to pulsate non-radially with several pulsation periods
which range between 8.4 hours and 1.77 days
(Lynas-Gray et~al. \cite{lks87}).  The spectral similarity indicates similar
temperatures, gravities, and chemical compositions.  The gravity of
LSS~5121 has not been determined, but it is clear from its spectrum that its
lies between those of the EHes \object{HD 160641} and \object{LS IV$+6\degr 2$}.
As BD$-9^\circ$4395,
LSS~5121 and HD~160641 all lie near or above the Z-bump instability boundary
(Saio \& Jeffery \cite{sj99}) it would be reasonable
to suppose that LSS~5121 may also pulsate. Given the similarity of the
physical properties of
LSS~5121 and HD~160641, one might also expect LSS~5121 to pulsate non-radially.
EHes with physical properties like those of LSS~5121 tend to
vary on timescales of hours to days, rather than minutes or months.  Our
observations were designed to detect variability on the expected timescales.

\object{LSS 5121} was too dim, $m_{\rm V} = 13.3$, for Lawson and Kilkenny
(\cite{lk98}) to test for photometric variations with the 0.5~m telescope used
in their observations. Although preliminary
observations of LSS~5121 at the Jacobus Kapteyn Telescope (JKT) in 1999 June
showed probable variability, LSS~5121 was not the primary target for the
observing run, so the observations did not cover enough time for the period
search to be conclusive (Aznar Cuadrado et al. \cite{arj2000}). The
observations we now report were carried out in 2000 May and June to provide 
better measurements of the star's variability.


\section{Observations and data reduction}

\subsection{JKT photometry}
Photometric measurements of \object{LSS 5121}
were performed using the JAG-CCD camera
on the 1-m JKT at the Roque de los Muchachos
Observatory, La Palma on the nights of 2000 May 13 and 14.  Observations
were made using the Cousins V filter.  Pixels on the $2048\times 2048$ chip
covered 0.33 arcsec per pixel, giving a usable field of about 10 arcmin.  Each
night the observations covered about 4.2 hours, with exposures
being taken every 3 or 4 minutes.

\begin{table}
 \begin{minipage}{70mm}
 \caption{Observations}
  \begin{tabular}{ll}
Telescope & ${\rm HJD - 2\,451\,000 } $ \\
\hline
JKT & 677.5626 -- 677.7374 \\
JKT & 678.5567 -- 677.7344 \\
McD 0.8-m & 706.8056 -- 706.9503 \\
McD 0.8-m & 707.9320 -- 707.9538 \\
McD 0.8-m & 709.7630 -- 709.9532 \\
McD 0.8-m & 710.7639 -- 710.9643 \\
McD 0.8-m & 711.7452 -- 711.9635 \\
McD 0.8-m & 717.7015 -- 717.9539 \\
McD 2.1-m & 709.7752 -- 709.9544 \\
McD 2.1-m & 710.9034 -- 710.9525 \\
McD 2.1-m & 711.7642 -- 711.9420 \\
 \end{tabular}
 \end{minipage}
\end{table}

\subsection{McDonald photometry}
\object{LSS 5121} was observed on the nights of 2000 June 10, 11, 13, 14, 15,
and 21 using
the Prime Focus Corrector of the McDonald Observatory 0.8-m telescope.
Observations were made using the Bessel V filter.  The camera gives 1.3
arcsec per pixel using the $2048\times 2048$ pixel CCD, giving a 46 arcminute
field.   As \object{LSS 5121} is in a fairly crowded field, we were
able to reduce the
portion of the chip read out, thus reducing the readout time and disk space
requirements, while still including enough comparison stars in the observation.
The photometric data from both McDonald and JKT were obtained for use as
differential photometry.
The times covered by observations at JKT and
McDonald are listed in Table~1.

\subsection{McDonald spectroscopy}
Spectra of \object{LSS 5121} were obtained on the nights of 2000 June 13, 14,
and 15 using the Sandiford Echelle spectrograph (McCarthy et~al.
\cite{mccarthy}) on the 2.1-m Otto Struve Telescope.  Repeated 15 to 20 minute
exposures were made during the times listed in Table~1 to allow us to detect
changes in the spectrum due to pulsation.  Spectra typically had a signal
to noise ratio between 7 and 11. The wavelength range of the echelle orders
overlapped, so the spectra give continuous coverage between 4410 and
4960~\AA .  We used a 1.65 arc sec slit width, which gives a spectral 
resolution of $\lambda / \Delta \lambda \approx 40\,000$.

\subsection{Data reduction}
Photometric data were reduced using standard {\sc iraf} packages to subtract
bias and divide by flats.  {\sc iraf}'s {\sc phot} package was used to do
aperture photometry on the images.  Stars in uncrowded parts of the field
and falling on parts of the CCD without bad pixels or columns were chosen
for use as comparison stars for the differential photometry
by visually inspecting the images.

Spectra were reduced using standard {\sc iraf} packages for bias and flat
field correction, reducing echelle orders to one dimensional spectra, and
applying the wavelength scale using thorium-argon arc spectra. Velocity
corrections for Earth's motion were found for each exposure using
{\sc rvcorrect} and were applied using {\sc dopcor}.

\section{Analysis}
\subsection{Differential photometry}
\begin{figure}
\resizebox{\hsize}{!}{\rotatebox{270}{\includegraphics{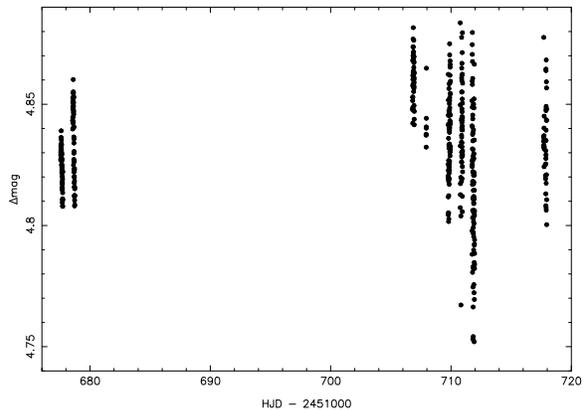}}} 
\caption{Differential photometry from the JKT and McDonald runs described in
Table~1.}
\label{MS1107f1}
\end{figure}

Comparison stars were chosen for each night's data.  To be chosen, a star needed
to be present on a good part of the chip for every \object{LSS 5121} image for
the night.  The flux measured for \object{LSS 5121} and the combined flux
measured for the comparison stars were used to find 
${\rm \Delta {\it mag} = {\it m}_V(LSS~5121) - {\it m}_V(comparison)}$.
Corrections were then made to put
all the data on the same scale, taking into account that different reference
stars were used on separate nights.  Differential photometry for the two
observing runs are shown in Fig.~\ref{MS1107f1}.

\subsection{Radial velocities}
\begin{figure}
\resizebox{\hsize}{!}{\rotatebox{270}{\includegraphics{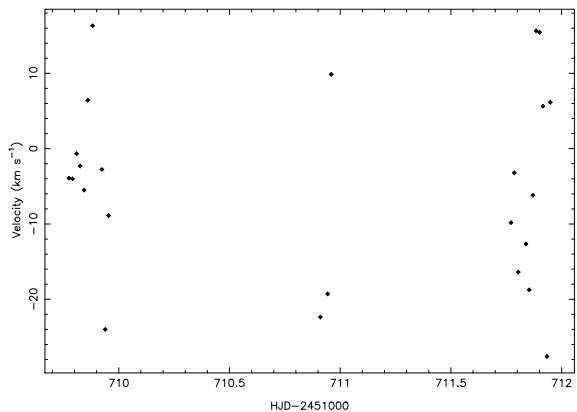}}}
\caption{Radial velocities measured for LSS~5121.}
\label{MS1107f2}
\end{figure}

Velocity shifts were measured using the cross correlation package {\sc fxcor}
in {\sc iraf}.  The spectra were coadded to provide a template with a signal
to noise ratio of about 35. Only the  two echelle orders covering 4460 to
4500~\AA\ and  4880 to 4930~\AA\ had features, notably the
 4471.5, 4920.8 and
4921.9~\AA\ \ion{He}{i} lines, strong enough for the cross correlation to
yield a velocity. We gave more weight to the velocities found using the
longer wavelength order, as the order containing the 4471.5~\AA\ line had lower
signal.  Other orders contained lines that may be useful for velocity
determination in studies using larger telescopes, but which did not give
reliable cross correlation peaks with our spectra.
The velocities thus found were applied
and the resulting spectra were coadded to give a template with smaller velocity
smearing for a second iteration.  Velocities found for three nights are
displayed in Fig.~\ref{MS1107f2}.
We estimate uncertainties in radial velocity determinations to be about
${\rm \pm 10\ km\ s^{-1}}$. This is a rough estimate based on how accurately
we could determine the center of the cross correlation peak.

\section{Results}
\begin{figure}
\resizebox{\hsize}{!}{\rotatebox{0}{\includegraphics{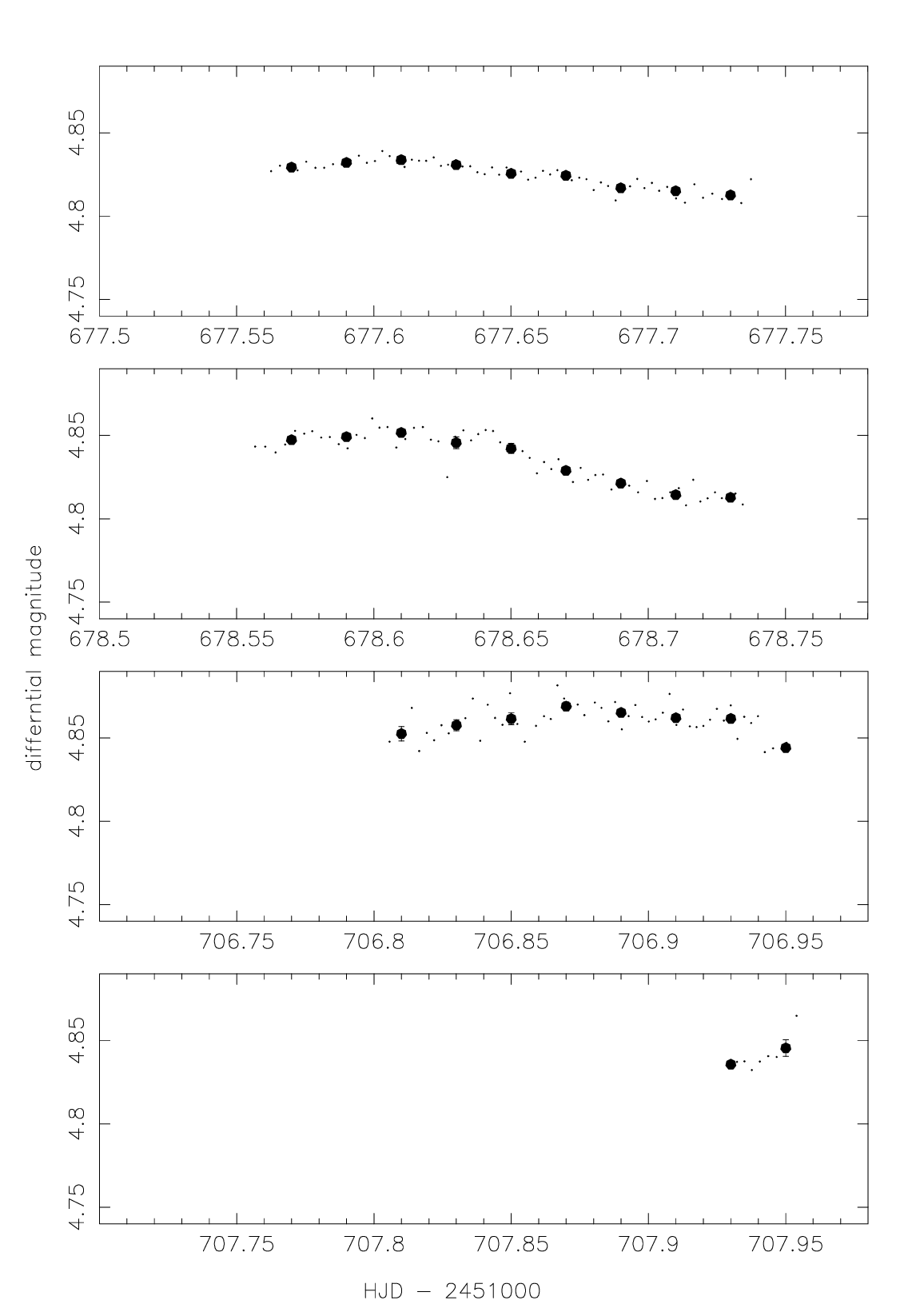}}} 
\caption{Differential photometry for \object{LSS 5121}.  Small points are 
measured photometry values.  Large points indicate average photometry in
0.02~day bins, with error bars showing standard deviation of the mean in each
bin. Top two panels are JKT data.  Bottom two are McDonald data. Error bars
are smaller than the large data points for the JKT data.}
\label{MS1107f3}
\end{figure}

\begin{figure}
\resizebox{\hsize}{!}{\rotatebox{0}{\includegraphics{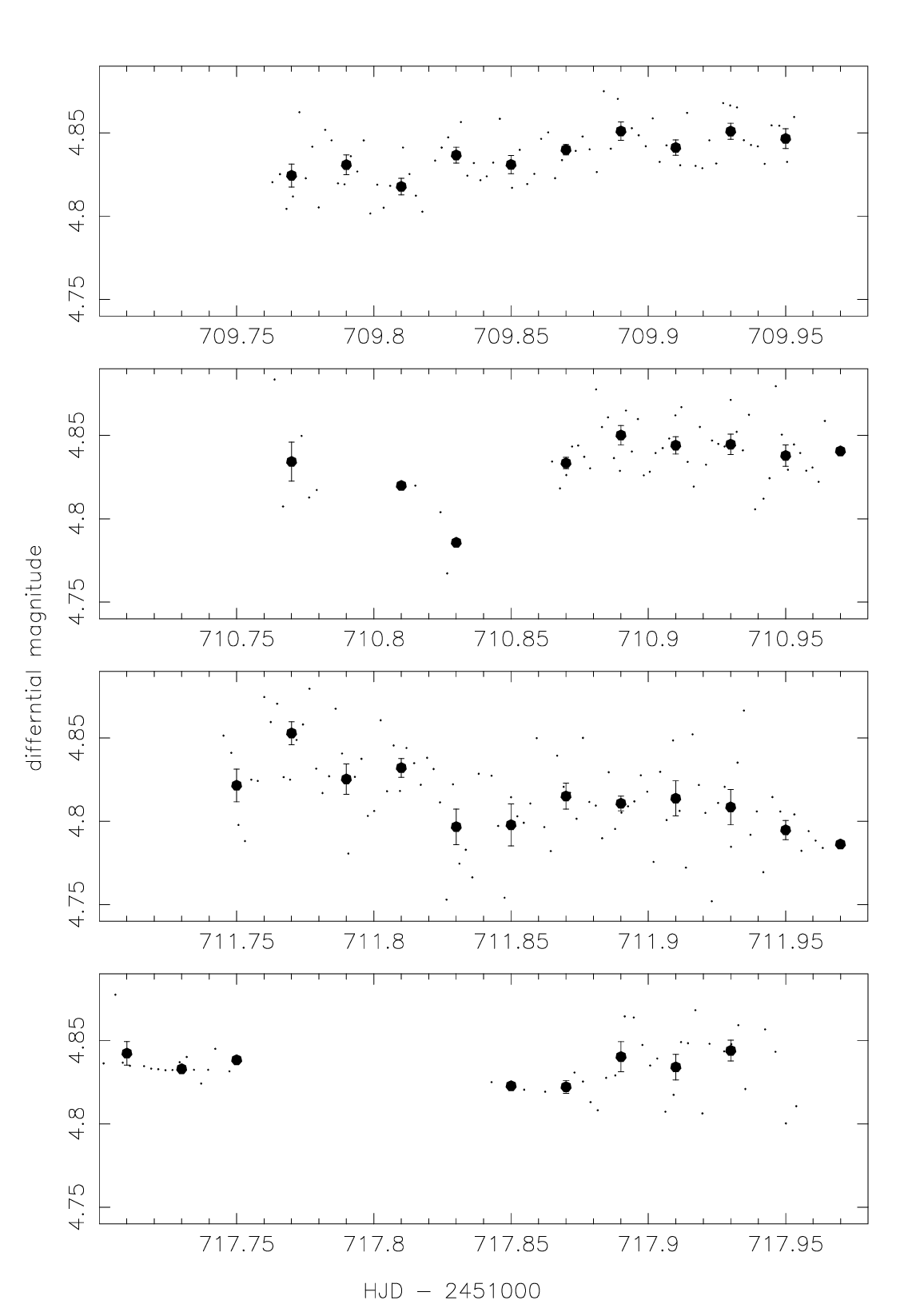}}} 
\caption{McDonald differential photometry.  Symbols as in Fig.~\ref{MS1107f3}.}
\label{MS1107f4}
\end{figure}

The uncertainties for the radial velocity data are
large compared to differences between the calculated velocities: we cannot
say that radial velocity variations are obvious.
The differential photometry shows evidence that
\object{LSS 5121} varies on a timescale of a few hours, however no period
was immediately apparent.  The photometry measured at JKT shows a magnitude
change of about 0.04~mag over a period of about 2.5 hours.
The photometry measured the next month at McDonald shows similar variation,
though the data are noisier.  Uncertainties for the photometric data
varied with the weather, but in general we estimate uncertainties of
$\pm 0.003$~mag for the JKT data and $\pm 0.015$~mag for the McD data.  Noise
was reduced by finding the mean for the photometric data in 0.02 day bins
(Figs.~\ref{MS1107f3} \& \ref{MS1107f4}).  Tests with non-variable comparison stars
with magnitudes similar to \object{LSS 5121} gave differential photometry which
did not vary more than about 0.01~mag.

\begin{figure}
\resizebox{\hsize}{!}{\rotatebox{270}{\includegraphics{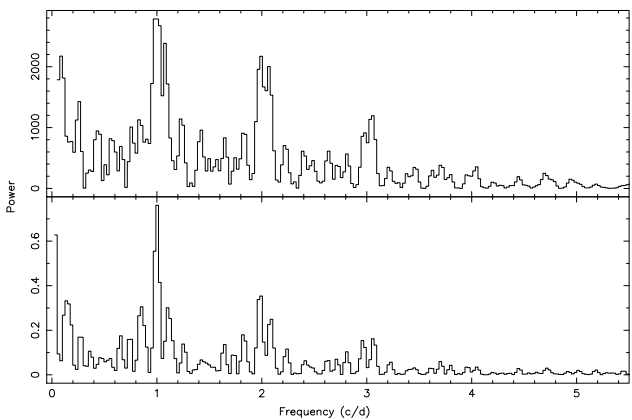}}} 
\caption{Periodogram (top panel) and window function (bottom panel)
for photometric data.}
\label{MS1107f5}
\end{figure}

Periodograms and window functions were found for the photometric and
velocity data using {\sc dipso}'s {\sc pdgram} and {\sc pdgwin} features
(Scargle \cite{sc82}).  The velocity periodogram has no major peaks, which is
not surprising, given the small number of data points.  The photometry
periodogram has no major peaks that are not also in the data's window function
(Fig.~\ref{MS1107f5}). The major peaks in both are near multiples of the one day
alias.
Data were phased to the periods indicated by minor peaks in the periodogram
not present in the window function, but this provided no further evidence of
periodic behavior.
We cannot rule out that LSS~5121 varies with a period of about 1 day or that
it varies with longer periods and/or smaller amplitudes
than our observations could detect.


\section{Discussion and Conclusions}

Differential photometry shows that \object{LSS 5121} varies in brightness, with
changes of up to 0.04~mag over a span of 2.5 hours. LSS~5121 was previously
known to have varying spectral lines.  We have now shown that it varies in
brightness as well.
Analysis of data from observations on 8 nights in 2000 May and June did not 
yield evidence of any periodicity in the variations. Our search for variability
in radial velocity was inconclusive.  A successful radial velocity search will
require a larger telescope to provide higher signal to noise spectra.

\begin{figure}
\resizebox{\hsize}{!}{\rotatebox{270}{\includegraphics{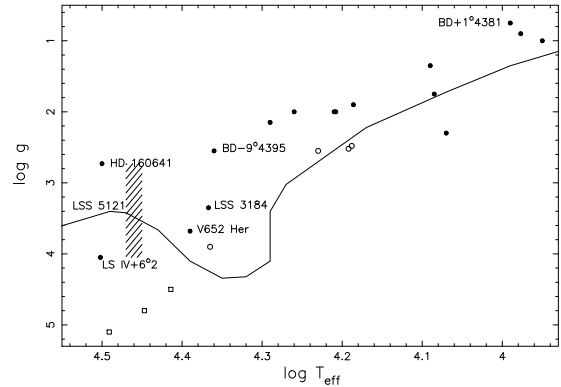}}}
\caption{Instability boundary (solid line) in the $\log g-\log T_{\rm eff}$
plane for 0.7~M$_\odot $ helium stars with $X=0.00$ and $Z=0.01$. EHes
are marked with circles.  The properties of the EHe LSS~5121 fall in the hatched
area.  Three He sdB stars are shown by squares.
Stars known to vary are indicated by filled symbols.}
\label{MS1107f6}
\end{figure}

Saio (\cite{s95}) showed that the region of the HR diagram where pulsations
should occur overlaps part of the region
where EHes appear. Figure~\ref{MS1107f6} shows the location of all known EHes with
known $T_{\rm eff}$ and $\log g$,
and the Fe-group opacity instability boundary for 0.7~M$_\odot $
helium stars in the $\log g-\log T_{\rm eff}$ plane.
The EHes mentioned in the text are labeled in the figure.
Helium stars above the boundary should be unstable against pulsation.
The exact location of the instability
boundary depends on stellar mass and chemical composition.

In general, the observed pulsation timescales in EHes decrease with higher
temperature (Saio \& Jeffery \cite{sj88}).  Two EHes, \object{LSS 3184}
(BX~Cir) and
\object{V652 Her} pulsate in their fundamental radial modes with well determined
periods of about 2.6 hours (Kilkenny et~al. \cite{klr96}, \cite{kkj99}).
Pulsations in the coolest EHes (e.g. \object{BD$+1\degr 4381$})
seem to be non-periodic or
chaotic with variation timescales of 8 to 25 days (Lawson et~al. \cite{lkw93}). 
Other EHes with known variability seem to pulsate non-radially,
some with multiple periods.
For example, Lynas-Gray et~al. (\cite{lks87}) report finding
periods of 0.35, 0.71, 1.12, and 1.77 days in the photometry of
\object{HD 160641}, an EHe with a temperature similar to that of
\object{LSS 5121} and Jeffery et~al. (\cite{jsh85}) report 3.5 day, 11.2 day,
and possibly additional periods for \object{BD$-9\degr 4395$}. It is possible
that chaotic pulsations also affect some of the hotter EHes, so that some
of the multiple periods found for the hot non-radial pulsators
would more accurately be referred to as variability timescales or
``quasi-periods,'' and that different
periods or no periods would be found if data from longer, more continuous
observations were available.  

Any pulsations of \object{LSS 5121} are either non-periodic or have
periods which  our data cannot resolve.  Our inability to
find a period is consistent with LSS~5121 being
a non-radial or a chaotic pulsator, though it does not constitute proof.
Its pulsations are probably similar to those of the hot
extreme helium stars \object{BD$-9\degr 4395$} and \object{HD 160641}, both of
which have been reported to pulsate non-radially with multiple periods
(Jeffery \& Heber \cite{jh92}, Lynas-Gray et~al. \cite{lks87}).

\begin{acknowledgements}
We acknowledge financial support from the UK PPARC (grant
Refs PPA/G/S/1998/00019 and PPA/G/O/1999/00058).
VW thanks Marcel Bergmann for instruction on use of the
0.8-m telescope and Nairn Baliber for a refresher on photometry reduction. GP
received support from grant AST 9618414 from the NSF.
\end{acknowledgements}

\end{document}